\documentclass[12pt]{article}
\usepackage{graphicx}
\usepackage{epsfig}
\usepackage{epstopdf}
\DeclareGraphicsExtensions{.pdf,.eps,.png,.jpg,.mps}

\usepackage{cite}

\def\lsim{\mathrel{\rlap{\lower3pt\hbox{\hskip0pt$\sim$}}
     \raise1pt\hbox{$<$}}}         %less than or approx. symbol
\def\gsim{\mathrel{\rlap{\lower4pt\hbox{\hskip1pt$\sim$}}
     \raise1pt\hbox{$>$}}}         %greater than or approx. symbol

\usepackage{amsmath}
\usepackage{amsfonts}

\begin{document}
\begin{titlepage}

%\centerline{\Large \bf Honey, Shrunk Sample Covariance Matrix is a Factor Model}
\centerline{\Large \bf Shrinkage = Factor Model}
\medskip

\centerline{Zura Kakushadze$^\S$$^\dag$\footnote{\, Zura Kakushadze, Ph.D., is the President of Quantigic$^\circledR$ Solutions LLC,
and a Full Professor at Free University of Tbilisi. Email: \tt zura@quantigic.com}}
\bigskip

\centerline{\em $^\S$ Quantigic$^\circledR$ Solutions LLC}
\centerline{\em 1127 High Ridge Road \#135, Stamford, CT 06905\,\,\footnote{\, DISCLAIMER: This address is used by the corresponding author for no
purpose other than to indicate his professional affiliation as is customary in
publications. In particular, the contents of this paper
are not intended as an investment, legal, tax or any other such advice,
and in no way represent views of Quantigic$^\circledR$ Solutions LLC,
the website \underline{www.quantigic.com} or any of their other affiliates.
}}
\centerline{\em $^\dag$ Free University of Tbilisi, Business School \& School of Physics}
\centerline{\em 240, David Agmashenebeli Alley, Tbilisi, 0159, Georgia}
\medskip
\centerline{(October 25, 2015)}

\bigskip
\medskip

\begin{abstract}
{}Shrunk sample covariance matrix is a factor model of a special form combining some (typically, style) risk factor(s) and principal components with a (block-)diagonal factor covariance matrix. As such, shrinkage, which essentially inherits out-of-sample instabilities of the sample covariance matrix, is not an alternative to multifactor risk models but one out of myriad possible regularization schemes. We give an example of a scheme designed to be less prone to said instabilities. We contextualize this within multifactor models.
\end{abstract}
\medskip
\end{titlepage}

\newpage

{}In his seminal work on mutual fund performance, Sharpe [1966] eloquently posits: ``The key element in the portfolio analyst's view of the world is his emphasis on both expected return and risk." Construction of a trading portfolio for equities schematically\footnote{\, Deliberately omitting many important details, that is.} can be thought of as consisting of two steps. First, one comes up with some expected returns for stocks in the trading universe. This drives the ``reward" part of the portfolio. Second, one constructs the portfolio holdings based on these expected returns. It is mainly this stage that deals with the ``risk" part.\footnote{\, Albeit, some elements of ``risk management" can be (and at times are) incorporated into the expected returns, e.g., sector or industry neutrality.}

{}Many incarnations of this second step, including mean-variance optimization \cite{Markowitz} and its numerous variations, Sharpe ratio \cite{Sharpe94} maximization, etc., require inverting a covariance matrix of returns. When the number of stocks in a portfolio is large and the number of available (relevant) observations in the historical time series of returns is limited, the sample covariance matrix (SCM) based on these historical returns is (badly) singular. Thus, if $M+1$ is the number of observations in the time series and $N$ is the number of stocks in the portfolio, SCM is singular if $M < N$.\footnote{\, Let $R_{is}$ be the time series of our returns, $i=1,\dots,N$, $s=0,1,\dots,M$. Let $X_{is} = R_{is} - \overline{R}_i$ be the serially demeaned returns, where $\overline{R}_i$ is the time-series mean of $R_{is}$. In matrix notation SCM is given by $C = {1\over M} X X^T$. (We are assuming $M\gg 1$, so the difference between the unbiased estimate with $M$ in the denominator vs. the maximum likelihood estimate with $M+1$ in the denominator is immaterial for our purposes.) Since at most $M$ columns of the matrix $X$ are linearly independent (as its column sums $\sum_{s=0}^M X_{is} = 0$), the rank of the matrix $C_{ij}$ is at most $M$ if $M < N$.\label{fn.SCM}}  Furthermore, unless $M\gg N$, which is rarely -- if ever -- the case in practice, the off-diagonal elements of SCM are not out-of-sample stable.\footnote{\, This statement is often regarded as stemming from empirical evidence. However, it is well-understood theoretically. We can always rotate our serially demeaned returns $X_{is}$ to an orthogonal basis and rescale them to have unit serial variances. Then the true covariance matrix is the $N\times N$ identity matrix. Pursuant to the Bai-Yin theorem \cite{BY}, the smallest and largest eigenvalues of SCM have the limits $\lambda_{min} = (1 - \sqrt{y})^2$ and $\lambda_{max} = (1 + \sqrt{y})^2$, where $y = N/M$ is fixed and $N,M \rightarrow\infty$. So for $M,N\gg 1$ we must have $M\gg N$ for all eigenvalues to be close to 1.\label{fn.BY}}

{}One method often used for mitigating these issues is the so-called shrinkage \cite{LW}. It is often regarded as an ``alternative" to multifactor risk models. However, as we discuss below, shrunk SCM is in fact a factor model of a special form based on a combination of some risk factors\footnote{\, These are typically, but not necessarily, style factors.} and principal components.

{}The idea behind shrinkage is simple. Instead of using SCM $C_{ij}$, one uses its weighted linear combination with another matrix (``shrinkage target"), call it $\Delta_{ij}$:
\begin{equation}\label{shrunk}
 {\widetilde C}_{ij} = q~\Delta_{ij} + \left(1 - q\right) C_{ij}
\end{equation}
Here the weight (``shrinkage constant") $0 \leq q \leq 1$. The matrix $\Delta_{ij}$ is assumed to be positive-definite and (relatively) stable out-of-sample. We must have ${\widetilde C}_{ii} = C_{ii}$, so $\Delta_{ii} = C_{ii}$. A priori $\Delta_{ij}$ can be otherwise arbitrary, in which case the ``shrunk" matrix ${\widetilde C}_{ij}$ can be thought of as a {\em regularization} of SCM $C_{ij}$. Thus, when $q \ll 1$, ${\widetilde C}$ approaches $C$. Furthermore, even when $C$ is singular, ${\widetilde C}$ is invertible for $q > 0$.

{}However, in practice $\Delta_{ij}$ must have some relevance to the underlying returns whose covariance matrix we are attempting to model. The simplest choice is a diagonal matrix $\Delta_{ij} = C_{ii}~\delta_{ij}$. A step up in complexity would be to use a 1-factor model.\footnote{\, In the model with uniform correlations used in \cite{LW} we have $\Delta_{ij} = \rho~\sigma_i\sigma_j$ for $i\neq j$, where $\sigma_i^2 = C_{ii}$. This is a special case of a 1-factor model $\Delta_{ij} = \xi_i^2~\delta_{ij} + \Omega_i~\Omega_j$ with $\xi_i^2 = \left(1-\rho\right)\sigma_i^2$, $\Omega_i^2 = \rho~\sigma_i^2$ and the $1\times 1$ factor covariance matrix absorbed into $\Omega_i$ (see below).} Let us be a bit more general here and take $\Delta_{ij}$ to be a $K$-factor model:\footnote{\, See, e.g., \cite{GK} and references therein.}
\begin{equation}
 \Delta_{ij} = \xi_i^2~\delta_{ij} + \sum_{A,B=1}^K \Omega_{iA}~\Phi_{AB}~\Omega_{jB}
\end{equation}
Here: $\xi_i$ is the specific (a.k.a. idiosyncratic) risk for each stock; $\Omega_{iA}$ is an $N\times K$ factor loadings matrix; and $\Phi_{AB}$ is a $K\times K$ factor covariance matrix (FCM), $A,B=1,\dots,K$. The number of factors $K\ll N$ to have FCM more stable than SCM.

{}We can now see that the ``shrunk" matrix ${\widetilde C}_{ij}$ is a factor model of a special form with a (block-)diagonal FCM. Thus, let us use the spectral representation:
\begin{equation}\label{spectral}
 C_{ij} = \sum_{a=1}^N \lambda^{(a)}~V^{(a)}_i~V^{(a)}_j
\end{equation}
where $\lambda^{(a)}$ are the eigenvalues of $C_{ij}$, and $V^{(a)}_i$ are the corresponding principal components (i.e, the eigenvectors normalized such that $\sum_{i=1}^N V^{(a)}_i~V^{(b)}_i = \delta_{ab}$), and the index $a=1,\dots,N$ is ordered such that $\lambda^{(1)} > \lambda^{(2)} > \dots > \lambda^{(N)}$. More precisely, we can have degenerate eigenvalues. For the sake of simplicity -- and this is not critical here -- let us assume that all positive eigenvalues are non-degenerate.\footnote{\, Also, we are assuming that there are no pairwise 100\% (anti-)correlated returns.} However, if $M < N$ (see above), we have null eigenvalues.\footnote{\, In practice these null eigenvalues can be distorted by computational rounding and turn into small positive or negative values. So, we assume that all such ``quasi-null" eigenvalues are rounded to 0. Furthermore, this assumes that there are no N/As in any of the time series of returns. If there are (non-uniform) N/As and SCM is computed by omitting such pair-wise N/As, then the resulting correlation matrix can have negative eigenvalues that are not ``small" in the above sense, i.e., they are not zeros distorted by computational rounding. In this case we can use the deformation method of \cite{RJ}. In any event, we assume that all $\lambda^{(a)}\geq 0$.} These null eigenvalues do not contribute to the sum in (\ref{spectral}) so  we can restrict it to the first $F$ values of $a$ for the positive eigenvalues. The ``shrunk" matrix ${\widetilde C}_{ij}$ can be written as a factor model:
\begin{equation}
 {\widetilde C}_{ij} = {\widetilde \xi}_i^2~\delta_{ij} + \sum_{\alpha,\beta=1}^{K+F} {\widetilde \Omega}_{i\alpha}~{\widetilde \Phi}_{\alpha\beta}~{\widetilde \Omega}_{j\beta}
\end{equation}
Here: the index $\alpha = (A, a)$ takes $K+F$ values; ${\widetilde \xi}_i^2 = q~\xi_i^2$; ${\widetilde \Omega}_{iA} = \Omega_{iA}$; ${\widetilde \Omega}_{ia} = V^{(a)}_i$, $a = 1,\dots, F$; ${\widetilde\Phi}_{AB} = q~\Phi_{AB}$; ${\widetilde \Phi}_{ab} = \left(1-q\right)\lambda^{(a)}\delta_{ab}$; and ${\widetilde\Phi}_{Aa} = 0$. So, we have a factor model with the $K$ factors from $\Delta_{ij}$ plus $F$ principal components (cf. \cite{MM}); however, $(K+F)\times(K+F)$ FCM ${\widetilde \Phi}_{\alpha\beta}$ is {\em ad hoc} set to be block-diagonal (it is diagonal for $K=1$), and its normalization relative to the specific risk (which is the rescaled specific risk from the $K$-factor model) is controlled by $q$.

{}Why is this observation useful? One evident issue with the ``shrunk" matrix ${\widetilde C}_{ij}$ is that it essentially inherits the out-of-sample instabilities of SCM as it uses all $F$ principal components with non-zero eigenvalues. A simple way of reducing this instability is to use fewer, first ${\widehat F} < F$ principal components. Consider the matrix
\begin{equation}\label{C.hat}
 {\widehat C}_{ij} = \nu_i~\nu_j~\Delta_{ij} + \sum_{a = 1}^{\widehat F}\lambda^{(a)}~V^{(a)}_i~V^{(a)}_j
\end{equation}
where, unlike in (\ref{shrunk}), there is no ``shrinkage constant" $q$ to determine (cf. \cite{LW}) as the coefficients $\nu_i$ are fixed from the requirement that ${\widehat C}_{ii} = C_{ii}$:
\begin{equation}\label{nu}
 \nu_i^2 = {1\over C_{ii}} \sum_{a = {\widehat F}+1}^F\lambda^{(a)}~\left[V^{(a)}_i\right]^2
\end{equation}
The matrix ${\widehat C}_{ij}$ too is a factor model with $K+{\widehat F}$ factors and a (block-)diagonal FCM. Equations (\ref{C.hat}) and (\ref{nu}) provide a simple {\em ad hoc} method for combining principal components with style, industry, etc. factors from a ``fundamental" factor model $\Delta_{ij}$, with the factor loadings $\Omega_{iA}$ for the $K$ factors rescaled by the coefficients $\nu_i$.

{}So, do all roads lead to Rome? Even shrinkage reduces to a factor model. To be precise, shrunk SCM is a factor model only if the matrix $\Delta_{ij}$ is a factor model. However, realistically, what else can it be in practice? If we knew how to write down a non-factor-model covariance matrix that approximates SCM well and is out-of-sample stable, we would not need shrinkage or anything else in the first instance!

{}Also, in shorter-horizon applications in many (if not most or even all) cases the number of available (relevant) observations $M < N$ (and often $M \ll N$). So, in this case SCM is singular (not just ``estimated with a lot of error" \cite{LW}). So shrinkage is essentially a regularization scheme. However, it is one out of myriad ways of regularizing SCM -- Equations (\ref{C.hat}) and (\ref{nu}) provide another such scheme with less inherent out-of-sample instability by design. Generally, higher principal components are not out-of-sample stable, the first principal component being most stable.\footnote{\, This can also be understood using the Bai-Yin theorem (see footnote \ref{fn.BY}): the limit for the largest eigenvalue $\lambda_{min} = (1 + \sqrt{y})^2$ holds even if $y = N/M > 1$. So, the first principal component is relatively stable, but higher ones are not as constrained and tend to be less stable.} This is one reason why statistical risk models based on principal components are not as popular commercially as ``fundamental" factor models.

{}The latter are based on style and industry risk factors. It is the (much more numerous than style) industry factors that provide not only higher granularity in ``fundamental" factor models, but also more stability, for a prosaic reason too -- stocks do not jump industries all that often! In fact, at shorter horizons one typically goes beyond the limited industry classification granularity employed by standardized commercial risk models and the number of industry factors can be in hundreds. This leads to an issue: sample FCM is singular when the number of observations is limited. Happily, a solution lies in using the industry classification hierarchy (e.g., ``sub-industries $\rightarrow$ industries $\rightarrow$ sectors" in the case of BICS) to sequentially reduce the size of the factor covariance matrix so it is computable \cite{Het}.

{}So, all roads do seem to lead to Rome... This should not come as a surprise. For $M < N$ SCM itself is nothing but an incomplete factor model (see footnote \ref{fn.SCM}):
\begin{equation}\label{SCM}
 C_{ij} = {1\over M} \sum_{s = 0}^M X_{is}~X_{js}
\end{equation}
It is missing the specific risk. Rotating and rescaling the demeaned returns $X_{is}$ and augmenting (\ref{SCM}) with the specific risk via a diagonal or factor-model $\Delta_{ij}$ results in (\ref{shrunk}). And this rotation is nothing but transforming to the principal component basis.


\begin{thebibliography}{99}

\makeatletter
\def\@biblabel#1{}
\makeatother

\bibitem[Bai and Yin, 1993]{BY} Bai, Z.D. and Yin, Y.Q.
``Limit of the smallest eigenvalue of a large dimensional sample covariance matrix."
The Annals of Probability, 21(3) (1993) 1275-1294.

\bibitem[Grinold and Kahn, 2000]{GK} Grinold, R.C. and Kahn, R.N.
``Active Portfolio Management." New York, NY: McGraw-Hill, 2000.

\bibitem[Kakushadze, 2015]{Het} Kakushadze, Z.
``Russian-Doll Risk Models." Journal of Asset Management, 16(3) (2015), pp. 170-185.

\bibitem[Ledoit and Wolf, 2004]{LW} Ledoit, O. and Wolf, M.
``Honey, I Shrunk the Sample Covariance Matrix."
The Journal of Portfolio Management, 30(4) (2004), pp. 110-119.

\bibitem[Markowitz, 1952]{Markowitz} Markowitz, H.
``Portfolio selection."
The Journal of Finance, 7(1) (1952), pp. 77-91.

\bibitem[Menchero and Mitra, 2008]{MM} Menchero, J. and Mitra, I.
``The Structure of Hybrid Factor Models."
Journal of Investment Management, 6(3) (2008) 35-47.

\bibitem[Rebonato and J\"ackel, 1999]{RJ} Rebonato, R. and J{\"a}ckel, P.
``The most general methodology to create a valid correlation matrix for risk management and option pricing purposes."
Working paper, 1999. Available online from: http://ssrn.com/abstract=1969689.

\bibitem[Sharpe, 1966]{Sharpe66} Sharpe, W.F.
``Mutual Fund Performance."
Journal of Business, 39(1) (1966), pp. 119-138.

\bibitem[Sharpe, 1994]{Sharpe94} Sharpe, W.F.
``The Sharpe Ratio."
The Journal of Portfolio Management, 21(1) (1994), pp. 49-58.



\end{thebibliography}
\end{document}